\documentstyle[12pt,epsfig]{article}
\newcommand{\be}{\begin{equation}}
\newcommand{\ee}{\end{equation}}

 1
\font\elevenrm=cmr10 scaled\magstep 1
 1

\def\ref{\hang\noindent}

\begin{document}
\vspace*{1.8cm}
  \centerline{\bf  HIGHLIGHTS from RXTE after 2.5 YEARS:}
  \centerline{\bf NEUTRON-STAR SPINS at KILOHERTZ FREQUENCIES,}
  \centerline{\bf MICROQUASARS and MORE}
\vspace{1cm}
  \centerline{HALE V. BRADT}
\vspace{1.4cm}
  \centerline{MASSACHUSETTS INSTITUTE OF TECHNOLOGY}
  \centerline{\elevenrm Room 37--587, Cambridge MA 02139--4307, USA}
  \centerline{\elevenrm E-mail: bradt@mit.edu}
\vspace{3cm}
\begin{abstract}
The \it Rossi X-ray Timing Explorer (RXTE) \rm satellite was launched on 30 December 
1995. It has made substantial contributions pertaining to compact objects and their 
environs. Broad-band spectral and short-time-scale temporal studies are exploring the 
effects of General Relativity in the regime of strong gravity. We present a brief outline of 
the principal contributions and then give a general overview of two
new areas of x-ray 
astronomy that have proven by RXTE to be very fruitful: accreting neutron stars with 
millisecond spin periods and microquasars. The former pertains to the spin evolution of 
low-mass x-ray binaries and the equations of state of neutron stars
while the latter is lends insight to disk-jet interactions in galactic black-hole 
binary systems.
\end{abstract}
\vspace{2cm}

\section{Mission status}

The \it Rossi X-ray Timing Explorer \rm (Bradt, Rothschild \& Swank 1993) was launched on 30 
December 1995. Since then it has carried out a diverse observing program that has been open to the entire astronomical community since a month after launch. It carries an All-Sky Monitor 
(Levine et al. 1996; Levine 1998) which, together with a flexible 
spacecraft pointing capability, permits rapid (hours) acquisition  of new or recurrent 
transient sources, sources entering new or interesting states, and gamma-ray burst 
afterglows. The pointed instruments are a large Proportional Counter Array (PCA; 2 -- 60 
keV; Jahoda et al. 1996) and a rocking High Energy X-ray Timing Experiment (HEXTE; 
15 -- 200 keV; Rothschild et al. 1998). All three instruments continue to operate close to 
their design state. It is hoped that operations can continue for several more years. There are 
no on-board expendables which would limit the spacecraft life.

Information about the mission as well as data products may be accessed on the web 
through: http://heasarc.gsfc.nasa.gov/docs/xte/. The source intensities from the 
ASM are posted every few hours on: http://heasarc.gsfc.nasa.gov/xte\_weather/.

\section{Scientific accomplishments: overview}

The \it RXTE \rm was designed to study compact objects and the material in their environs with 
emphasis on temporal studies with high statistics together with broad
band-band spectroscopy. Over 150 papers 
had been accepted in the refereed literature by the summer of 1998. \it RXTE \rm has been highly 
influential in conduct of science in many wavebands. Over 120 IAU Circulars had 
announced \it RXTE \rm discoveries of immediate interest and these have been followed by 
numerous reports from other observatories, gamma-ray, x-ray, radio and optical. At 
http://heasarc.gsfc.nasa.gov/whatsnew/xte/papers.html, circulars and papers may be found.

The areas in which \it RXTE \rm has made important contributions are listed here, with a few 
sample references. They are extracted from the Proposal to the 1998 Senior Review of NASA 
Astrophysics Missions Operations and Data Analysis authored by J. Swank,
F. Marshall \& the \it RXTE \rm Users'
Group. Thereafter, I will give brief overviews, from my perspective, of two areas wherein \it RXTE \rm has broken substantial new ground, namely, kiloHertz oscillations and microquasars.

1. Behavior of matter in regimes of strong gravity through the temporal and spectral 
signatures of kiloHertz pulsars and variability in microquasars (see refs. below). 

2. Spinup evolution of neutron stars through the characteristics of kHz pulsars and the 
discovery of the first accretion powered millisecond pulsar (see refs. below).

3. Formation of relativistic astrophysical jets through the multiwavelength study of 
(galactic) microquasars (see refs. below).

4. AGN unified models and emission mechanisms through multiwavelength (esp. TeV) 
studies (\it e.g., \rm Cantanese et al. 1997), detection of iron line and reflection components 
in individual Sy1 and Sy2 galaxies (\it e.g., \rm Weaver, Krolik \&
Pier 1998, Nandra et al. 
1998), and temporal studies with long term sampling with both the PCA and the ASM 
instruments. See for example the variability of the BL Lac objects Mkn 501 and Mkn 421 
in Fig. 1. 

5. High magnetic fields in neutron stars through the study of (1) the magnetosphere/disk 
boundary with low-frequency QPOs (\it e.g., \rm Kommers, Chakrabarty \& Lewin 1998), 
Type II bursts, the propeller effect (\it e.g., \rm Cui 1997), and cyclotron lines 
(Kreykenbohm et al. 1999); (2) the discovery of the fastest rotation powered pulsar ($P = 
16$ ms; Marshall et al. 1998), and (3) the discovery of x-ray pulsations supporting the 
identification of a ÔÔmagnetarÕÕ, a neutron star with an 
extraordinarily high magnetic field ($2 \times 10^{10}$ T), as a soft gamma-ray repeater 
(Kouveliotou et al. 1998).

6. Transient sources through PCA slews and ASM monitoring which have revealed 
$\sim$15 previously unknown sources and numerous recoveries of previously known 
sources, together with follow-on studies with the PCA and other observatories. Several new 
examples of radio-jet systems have been revealed (see below). Sample light curves from the RXTE/ASM extending over 2.5--yr are 
shown in Fig. 1.

7. State changes in binary systems through temporal/spectral tracking during major changes 
in x-ray flux and spectrum. In the case of Cyg X--1, a change of corona size is indicated 
(Cui et al. 1997). 

8. Superorbital quasi periodicities in high and low mass binaries as well as in black-hole 
binaries through their discovery or confirmation with the ASM. (\it e.g., \rm 
$P \approx $ 60 d in SMC X1; See Fig. 1). Some are most likely due to precessing accretion disks 
similar to the 35--d period in the well known Her X--1. However, the evolution of wave 
forms and periods indicates relatively complex underlying physics (Levine 1998).

9. Gamma-ray burst afterglows through rapid position determinations with the ASM and 
PCA. Five burst positions with positions accurate to a few arcminutes in one or two 
dimensions have been reported to the community within hours of the event (Smith et al. 
1999). One of the three known GRB with measured extragalactic red shifts (GRB 
980703) was first located with the ASM on \it RXTE \rm (Levine, Morgan \& Muno 1998, 
Djorgovski et al. 1998). Another (GRB 970828) had a very bright afterglow in x rays but no 
discernable optical or radio afterglow (Remillard et al. 1997; Groot et al. 1998).

10. X-ray emission regions in cataclysmic variables through PCA tracking of eclipse 
transitions with precisions of tens of kilometers (\it e.g., \rm Hellier 1997), and wind-wind 
collisions in Eta Carina through repeated PCA spectral observations (Corcoran et al. 1997). 

11. Diffuse source spectra from the galactic plane (Valinia \& Marshall 1998), supernova 
remnants (\it e.g., \rm Allen et al. 1997), and clusters of galaxies (\it e.g., \rm Rephaeli \& 
Gruber 1999) to high energies ($>$ 10 keV) with PCA and HEXTE.

\section {KiloHertz oscillations in low-mass x-ray binaries}

The most prominent area of \it RXTE \rm accomplishment is that of
kiloHertz oscillations. Relatively high-Q quasiperiodic oscillations (QPO) at kHz 
frequencies (up to 1230 Hz) in Low-Mass X-ray Binaries (LMXB) have been found 
in the persistent flux of 18 sources (as of this writing, Dec. 1998). Five of these, and one other,
exhibit quite coherent, but transient, oscillations in the frequency range 290 -- 590 Hz during Type I (thermonuclear) bursts. One additional source, a transient,
exhibited sustained coherent pulsations at 401 Hz with Doppler shifts
characteristic of a binary orbit. Reviews of the field may be found in
van der Klis (1998, 1999).

These new phenomena are probing the processes taking place close to the 
neutron stars where the effects of General Relativity are important.
For example, if the highest-frequency kHz QPO observed in a 
given source is interpreted as the Kepler frequency of the inner accretion disk, it should be 
limited by the frequency of the innermost stable orbit allowed in
general relativity ($r=6GM/c^2$ 
for Schwarzschild geometry). Since 17 of the sources exhibit maximum 
frequencies in the relatively narrow 1000--1200 Hz range, these frequencies may indeed 
represent the innermost stable orbits. The phenomenon is also placing
constraints on the equations of state of neutron stars as we
illustrate below.

The first discovered examples of the quasi-periodic oscillations at kHz
frequencies were in Sco X-1 (van der Klis et al. 1996) 
and 4U 1728--34 (Strohmayer et al. 1996). These QPO usually 
occur in pairs. As the source intensity increases, the two QPO generally increase in frequency with 
a frequency difference that remains approximately constant (Strohmayer et al. 1996); see Fig. 
2.

If the higher-frequency peak of the pair is the Kepler velocity of a blob of orbiting disk 
material, the lower-frequency peak could arise from the interaction of the blob with the  
magnetosphere which is co-rotating with the spinning neutron
star. The observed lower frequency would thus be a beat frequency, such as that postulated for 
QPOs at much lower frequencies ($\le 60$ Hz) in the 1980's (Alpar and Shaham 1985, Lamb
et al. 1985).  The difference of the two frequencies would be the neutron-star spin frequency. For the 17 sources that exhibit 
two frequencies, the differences range from $\sim$250 to $\sim$350 Hz,
indicating neutron-star spins in this range. (See Table 1 in van der
Klis 1999.) 

The 
increase of frequency of the oscillations with intensity (Fig. 2) can be understood in this picture as 
being due to an increase in the Kepler frequency arising from a
decrease in the size of the magnetosphere caused, in turn, by the increased ram 
pressure of the accreting material (Ghosh \& Lamb 1992). In fact, the source 4U 1820--30 exhibits frequencies 
that increase with flux until they saturate at about 1060 Hz and 800
Hz (Fig. 3). This suggests that the innermost stable orbit 
has been reached (Zhang et al. 1998b). But since this plot is a compilation of data from 
different observations that might have differing intensity-frequency
relations (see below), the effect 
could be an artifact (Mendez et al. 1999).

If indeed the maximal frequencies represent the innermost stable
orbits, the highest observed frequency seen to date (1228 Hz) yields a
neutron star mass of $\sim$2.0 $M_{\odot}$, which is significantly above
the canonical 1.4 $M_{\odot}$. Further, the radius of the neutron star
must not exceed the radius of the Kepler orbit corresponding to the
maximum observed frequency in a given source. This limit, together
with the marginally stable orbit just discussed above, places
constraints on allowed equations of state for neutron stars (Miller,
Lamb \& Psaltis 1998, see Fig. 4). The current limits do not yet
distinguish among the plotted equations of state, but the potential
for doing so is
clearly there. 

The interpretation that the difference frequency is the neutron-star
spin frequency gains credence from the discovery of nearly coherent
pulsing during x-ray bursts in several 
sources at about, or at about twice, the difference 
frequency (\it e.g., \rm Strohmayer et al. 1996). The frequency of the
burst oscillations in 4U 1728--34 is stable from burst to burst within
about 0.01\% over a period of 1.6 years 
(Strohmayer et al. 1998; Fig. 5). This stability is a strong indicator that these pulsations directly 
represent the 
neutron-star spin. They could arise from a transient hot spot (or hot spots) in the runaway 
thermonuclear burning on the neutron-star surface (Bildsten 1995). If there were a hot spot 
at each of two opposed magnetic poles, the detected frequency would be twice the spin 
frequency. 

This picture needs refinement or modification for several reasons: (1) the frequency 
difference as a function of intensity (or of one of the two frequencies) is not constant in Sco X--1
(van der Klis et al. 1997), 4U 1608 (Mendez et al. 1998), 4U 1735--44
(Ford et al. 1998), and possibly in all 
sources (Psaltis et al. 1998), (2) the frequencies in bursts are not strictly 1.0 or 2.0 times 
the difference frequencies in all sources, especially in 4U
1636--536 (Mendez, van der Klis \& van Paradijs 1998), (3) there are small frequency drifts during a 
single burst (Fig 5), and (4) although there are correlations between source intensity and 
QPO frequency on short time scales (hours), the correlations do not hold up over long 
periods (days); very different intensities can yield the same QPO
frequency, \it e.g., \rm in Aql X--1 (Zhang et al. 1998a). These 
problems are not necessarily fatal to the basic picture; there are
various proposed scenarios to 
explain the discrepancies. Alternatively, the answers could lie in
very different directions, see, \it e.g., \rm Stella \& Vietri (1999).

The so-called discrepancies are actually excellent probes with which to verify or discard theoretical 
models. For example, the frequency difference as a function of intensity in Sco X--1 requires 
quantitative understanding; see, for example, Stella \& Vietri
1999. Also, the frequency \it vs \rm
intensity dilemma (item 4 above) has been clarified by the discovery of monotonic mapping of frequency with 
position in the color-color diagram of 4U 1608--52 (Mendez et al. 1999). 

Finally, as noted above, the beat-frequency model was originally
introduced to explain some of the 
lower-frequency quasi-periodic oscillations ($\sim$ 6 -- 60 Hz) in LMXB in the 
1980's. One cannot explain \it both \rm kinds of oscillations (low-frequency and kHz) in the same 
source with this one model. Attempts to rationalize these phenomena include (1) a sonic-point model to explain the kHz oscillations as arising from interactions between radiation and 
orbiting blobs at the sonic-point radius (Miller, Lamb \& Psaltis
1998), (2) nodal precession of the inner disk, dominated by the
Lense-Thirring effect, to explain the lower-frequency oscillations (Stella \& 
Vietri 1998), and (3) periastron precession to explain the lower-frequency peak of the kHz 
twin peaks (Stella \& Vietri 1999). The latter model can reproduce
the changes in the frequency 
difference but does not attempt to explain the apparent coincidences with the frequencies of the kHz 
QPO during bursts in some sources. Strong-field GR is required to calculate the periastron-precession 
frequency. Thus, as pointed out by the authors, if their model is
validated, the kHz QPO phenomenon provides an unprecedented 
testbed for strong-field General Relativity.

These indicators of the neutron-star spin are only indirect (through the beat frequency) or 
fleeting (during bursts). The detection of coherent, persistent,
accretion-powered pulsing at millisecond periods 
had so far eluded \it RXTE \rm researchers. This elusive goal was reached with the recent (April 
1998) \it RXTE \rm discovery of highly coherent 401--Hz x-ray 
pulsing in the persistent flux of a transient source (Wijnands \& van der Klis 1998). A 
binary orbit was easily tracked with the
Doppler shifts of the 401--Hz pulsations (Chakrabarty \& Morgan 1998, Fig. 6). The 
orbital period is 2.01 hr which indicates a companion mass less than 0.1 $M_{\odot}$. 

The Doppler variation demonstrated without doubt that the pulsations arose from the neutron-star 
spin. The source thus became the first known \it accretion-powered
millisecond pulsar.\rm  
The source had been detected in a previous transient episode
(September 1996) with the
SAX Wide Field Camera (and also in RXTE/ASM data retrospectively) and two
x-ray bursts were observed from it (in 't Zand et al. 1998). It is
known as SAX J1808.4--3658. The rapid pulsing discovery occurred
during a later outburst (April 1998) that was revealed in RXTE/PCA data
during a spacecraft slew (Marshall 1998).

These discoveries of millisecond-period x-ray sources fill an 
important link in the spin evolution of neutron stars. It had long been postulated that 
millisecond radio pulsars are spun up in x-ray binaries (Radhakrishnan \& Srinivasan 
1982, Alpar et al. 1982), and LMXB were prime candidates because of their lack of 
coherent pulsations at lower frequencies. This long-sought evolutionary link has now been 
established. This is the successful attainment of one of \it RXTE's \rm major goals.  

\section{Microquasars}
Another area of major  accomplishment by \it RXTE \rm is that of
``microquasars''. The discovery of transient galactic x-ray emitting
objects with superluminal radio jets, GRO 1655--40 and GRS 1915+105 (\it e.g., \rm Tingay et al. 1995, Mirabel \& Rodriguez 1994), 
and the well-determined high mass ($7.0 \pm 0.2$ $M_{\odot}$) of the compact object in 
GRO 1655--40 (Orosz \& Bailyn 1997) focused attention on the fact that counterparts of 
(black-hole) quasars are close by in the Galaxy. Their proximity allows studies with much 
higher statistics, and their lower (stellar) masses lead to much smaller time constants for 
motions of matter in the vicinity of the compact object. The time constants 
scale linearly with mass, \it e.g., \rm the orbital period of Kepler matter in the innermost 
stable orbit. Thus a 1--year intensity variation in the vicinity of a $10^8$--$M_{\odot}$ quasar 
would occur in 3 s in a galactic 10--$M_{\odot}$ microquasar. 

Other galactic x-ray sources are known to exhibit evidence for radio jets through 
episodic non-thermal radio emission and/or diffuse emission or
resolved jets. These include the long-known Cyg X--3, GX
339--4, Cir X--1, SS433 and Cyg 
X--1 (see van Paradijs 1995 for references), and also the \it RXTE \rm discovered or recovered 
transients XTE J1748--288 (Rupen \& Hjellming 1998), GRS 1739--278 (Hjellming et al. 1996, 
Durouchoux et al 1996) and CI Cam (Hjellming \& Mioduszcwski 1998). One of these 
sources, Cir X--1 most likely contains a neutron star (Tennant, Fabian \& Shafer 1986, 
Shirey, Bradt \& Levine 1999), and another, CI Cam, is a symbiotic system. Altogether 
these sources are a rich resource for the understanding of the role accretion disks play in jet 
formation.

It is fortunate that the superluminal sources GRS 1915+105 and GRO 1655--40 have 
exhibited extensive activity during the \it RXTE \rm mission. GRO 1655--40 was active for about 
16 months beginning in April 1996, and GRS 1915+105 has been active since the beginning of the mission. The 
latter source exhibits a variety of states in its long-term variability as 
measured with the RXTE/ASM (Fig. 7). In the high-statistics data from the RXTE/PCA, 
its x-ray variability is dramatic and varied, including rapid oscillations, 
sudden dips, sharp spikes, etc., all accompanied with spectral changes (Fig. 8; Greiner, 
Morgan \& Remillard 1996, Morgan, Remillard \& Greiner 1997, Taam, Chen \& Swank 
1997). 

Next, I present briefly three areas of substantive progress in
microquasar studies.

\subsection{Initiation of accretion}

The sudden turn-on of GRO 1655--40 in x rays was fortuitously
monitored in BVRI during the week just prior to the x-ray turn on (Fig. 9; Orosz et al. 1997). A linear increase of flux 
was seen in all four bands. It began first in the I band, 6.1 d before the commencement of 
the linear x-ray rise. The increases in the R, V and B bands commenced systematically later with the 
latter occurring 5.0 days before the x-ray commencement. This sequence
suggests that the initiating event of a transient outburst was  a wave 
of instability propagating inward in the disk 
(Lasota, Narayan \& Yi 1996). This is an important breakthrough in the determination of 
the causes of x-ray nova outbursts.

\subsection{Accretion-jet correlations}

There have been clear coincidences between radio/infrared non-thermal flares and x-ray 
events, both on the longer time scales of the ASM data (Pooley \& Fender 1997) and shorter-term 
events in the PCA data (Pooley \& Fender 1997, Eikenberry et al. 1998, Mirabel et al. 1998). The latter type of x-ray event 
consists of a large x-ray dip ($\sim$15 minutes) that contains a pronounced spike (Fig. 8, bottom 
panel). Such an event is associated with an infrared flare and a delayed radio flare as shown in an 
event captured by Mirabel et al. (1998, Fig. 10). Five and possibly six IR/x-ray coincidences of this type 
were reported by Eikenberry et al. (1998) and in no case was the coincidence violated! 

These x-ray dips are repetitive, occurring irregularly at intervals of a half hour or so (Fig. 8). The 
source may reside in this 
state for hours to days. This is only one of several oscillatory states in which the 
source can find itself; see Fig. 8. The spectral evolution of an infrared/radio flare has been shown to represent a single 
relativistically expanding plasmoid (Eikenberry \& Fazio 1997, Mirabel et al. 1998, Fender \& Pooley 1998). These IR/radio events are small, \it i.e., 
\rm mini flares. 
A series of them emitted when the source is in this state could give rise to a single large 
superluminal outburst. It thus appears that the jets are quantized,
not continuous, and that \it RXTE \rm is seeing the ``pump'' that creates them!

X-ray spectral fits (Fig. 11, Swank et al. 1997) during these events show a softening of the 
disk-black body component, which can be interpreted as the disappearance of the inner part 
of the disk as proposed by Belloni et al. (1997a,b). Thereafter, the 
gradually increasing temperature and decreasing radius of the disk component would 
represent the refilling of the disk. The power-law component suddenly softens at a sharp 
x-ray spike near time 1600 s when the disk is nearly full. Mirabel et al. (1998) suggest that  
this spike is the initiating event of the flare (see the IR flare in Fig. 10). The frequency of the associated low-
frequency QPO (Fig. 11) appears qualitatively to track the disk radius as if it were the Kepler 
frequency at this or an associated radius. But the situation is not
this simple given the existence of other 
QPOs, \it e.g., \rm 67 Hz, in the system (see Remillard et
al. 1999).

As noted, GRS 1915+105 exhibits some half dozen states with different temporal/spectral 
variability, not all of which fit this simple disk-depletion picture. Additional multifrequency 
studies are needed as are more comprehensive models.  

\subsection{High-frequency QPO in Microquasars}

The microquasars exhibit quasi periodic oscillations (QPO) that are quite variable in 
frequency and also some that are relatively stable. These QPO have a large potential for 
probing the physics of the systems. The highest frequencies (Fig. 12), namely 67 Hz in GRS 
1915+105 (Morgan, Remillard \& Greiner 1997) and 300 Hz in GRO 1655--40 (Remillard et al. 1999) do not drift in frequency. They have led to intriguing speculation about 
their origins. The high frequencies place them close to the central black hole, and models 
usually invoke General Relativity. Suggested origins include the innermost stable orbit 
(Morgan et al. 1987), Lense-Thirring precession (Cui, Zhang \& Chen 1998), diskoseismic 
oscillations (Nowak, et al. 1997), and oscillations in the centrifugal barrier (Titarchuk, 
Lapidus, Muslimov 1998).

Some investigators are using these data and models to arrive at the angular momentum of 
the central black hole. The black-hole mass, $7.0 \pm 0.2$
$M_{\odot}$, of GRO 1655--40 and the 300--Hz
oscillations in this source suggest negligible black-hole angular 
momentum \it if \rm the oscillations are the Kepler frequency of the innermost stable orbit. On the 
other hand, if the 300 Hz oscillations are due to Lense-Thirring precession in the inner 
disk, they imply a maximally rotating black hole (Cui, Zhang \& Chen 1998). The latter view gains 
some support from the measured high disk temperature which is indicative of the small 
inner disk radius expected for prograde orbital motion of a maximally rotating black hole 
(Zhang, Cui \& Chen 1997). These conclusions are highly model dependent and therefore 
uncertain. Nevertheless, it is impressive to that the angular momentum of black holes 
is now being addressed by the community with data from \it RXTE \rm. This was not 
dreamed of even a few years ago.

All in all, it is clear that 
the jet formation processes, the conditions of disk stability, and the
formation of the power-law 
component are being explored with a powerful and effective tool, namely 
the temporal/spectral/statistical power of \it RXTE\rm. The behavioral
detail now being acquired from microquasars extends well 
beyond that which can be obtained from the much more distant
extragalactic quasars.

\section{Conclusions}

The \it RXTE \rm is making important strides in the study of compact objects, both galactic and 
extragalactic, in a wide variety of studies by a large international community of observers. 

The discovery of 401--Hz kHz coherent pulsations in a low-mass x-ray binary has 
established definitively an important link in the evolution of neutron
stars. The kHz QPO in 18  systems and coherent pulsations during bursts give additional strong indications of 
neutron-star spins at frequencies of a few hundred Hz. These QPO provide information about the 
behavior of matter in the immediate vicinity of the neutron star and are placing limits on the 
possible equations of state of neutron stars. 

The temporal/spectral signatures of the various behaviors in
microquasars are diverse, yet repeatable and well 
described with high statistics. They are powerful probes of these
systems and should serve as powerful discriminators of 
models. At the same time, the complexity makes difficult the construction of a 
comprehensive model of the emission processes. The results currently
point toward black-hole masses and angular momenta, the nature of disk
instabilities, and the precise events that initiate the jets signified
by radio/IR flares. These results clearly have applicability to
extragalactic quasars.

The temporal variability of x-ray spectra 
can, in principle, track the changing geometry of the several physical components of the 
system (\it e.g., \rm disk and corona). However this requires that these physical 
components be securely identified with the spectral components. This
is a major challenge now confronting microquasar researchers.

\it RXTE \rm studies are probing phenomena where strong General Relativity is important 
because of the proximity of the emitting plasmas to the central gravitational object. For 
example, frame dragging has been invoked for some high frequency QPOs, and the orbital 
frequency at the innermost stable orbit may have been encountered in LMXB systems. 
Measurements of these and other GR effects are now within the realm of \it RXTE \rm 
capabilities.

\section*{Acknowledgments}

The author is grateful for the efforts of the entire \it RXTE \rm team and the many 
observers whose work has contributed to the productivity of \it RXTE
\rm. He is especially grateful to the 
staff and students of the \it RXTE \rm group at M.I.T. for many
helpful and stimulating conversations. Helpful comments for this
manuscript were provided by R. Remillard and L. Stella.
This work was supported in part by NASA under contract
NAS5--30612. The author further acknowledges with gratitude the
support and hospitality provided to him during his sabbatical year at
the Osservatorio Astronomico di Roma. This report was completed while
overlooking the ``Pines of Rome''. 

\section{References}

\noindent Allen, G. E. et al. 1997, ApJ, 487, L97

\noindent Alpar, M. \& Shaham, J. 1985, Nature, 316, 239

\noindent Alpar, M. Cheng, A., Ruderman, M. \& Shaham, J. 1982, Nature, 300, 728

\noindent Belloni, T., Mendez, M., King, A. R., van der Klis, M. \& van Paradijs, J. 
1997a, ApJ, 479, L145

\noindent Belloni, T., Mendez, M., King, A. R., van der Klis, M. \& van Paradijs, J. 
1997b, ApJ, 488, L109

\noindent Bildsten, L. 1995, ApJ, 438, 852

\noindent Bradt, H. V., Rothschild, R. E. \& Swank, J. H. 1993, A\&AS, 97, 355

\noindent Cantanese, M. et al. 1997, ApJ, 487, L143

\noindent Chakrabarty, D. \& Morgan, E. 1998 Nature, 394, 346

\noindent Corcoran, M. F., Ishibashi, K., Swank, J., Davidson, K., Petre, R. \& Schmitt, 
M. 1997, Nature, 390, 587

\noindent Cui, W. 1997, ApJ, 482, L163

\noindent Cui, W., Zhang, S. N., Focke, W. \& Swank, J. H. 1997, ApJ, 484, 383

\noindent Cui, W., Zhang, S, N. \& Chen, W. 1998, ApJ, 492, 53

\noindent Djorgovski, S. G. et al. 1998, ApJ, 508, L17

\noindent Durouchoux P. et al. 1996, IAU Circ. 6383

\noindent Eikenberry, S.S. \& Fazio, G. G. 1997, ApJ, 475, L53

\noindent Eikenberry S. S., Matthews, K., Morgan, E. H., Remillard, R. A. \& Nelson, 
R. W. 1998, ApJ, 494, L61

\noindent Fender, G. G. \& Pooley, R. P. 1998, MNRAS 300, 573 

\noindent Ghosh, P. \& Lamb, F. K. 1992, in X-ray Binaries \& Recycled Pulsars, ed. E. 
P. J. van den Heuvel \& S. A. Rappaport (Dordrecht: Kluwer) 487

\noindent Ford, E. C., van der Klis, M., van Paradijs, J., Mendez, M.,
Wijands, R. \& Kaaret, P. 1998, ApJ, 508, L155

\noindent Greiner, J., Morgan, E. H. \& Remillard, R. A. 1996, ApJ, 473, L107

\noindent Groot, P. J., et al. 1998, ApJ, 493, L27

\noindent Hellier, C. 1997, MNRAS, 291, 71

\noindent Hjellming, R. M. 1996, IAU Circ. 6383

\noindent Hjellming, R. M. \& Mioduszcwski, A. J. 1998, IAU Circs. 6857, 6862, 6872

\noindent in 't Zand, J. J., Heise, J., Muller, J. M., Bazzano, A.,
Cocchi, M., Natalucci, L. \& Ubertini, P. 1998, Astr. Astrophys., 331, L25

\noindent Jahoda, K. et al. 1996, in EUV, X-ray, and Gamma-ray Instrumentation for 
Space Astronomy VII, ed. O. H. W. Sigmund \& M. A. Grummin, Proc. SPIE 2808, 59

\noindent Kommers, J. M., Chakrabarty, D. \& Lewin, W. H. G. 1998, ApJ, 497, L33

\noindent Kouveliotou, C., et al. 1998, Nature, 393, 235

\noindent Kreykenbohm, I., et al. 1999, A\&A, submitted (astro-ph 9810282)

\noindent Lasota, J. P., Narayan, R. \& Yi, I. 1996, A\&A, 314, 813

\noindent Lamb, F. K., Shibazaki, N., Alpar, M. \& Shaham, J. 1985, Nature, 317, 681

\noindent Levine, A. M. 1998, Nucl. Phys. B (Proc. Suppl.), 69/1--3, 196 [Proc. of ÔÔThe Active 
X-ray SkyÕÕ, eds. L. Scarsi, H. Bradt, P. Giommi \& F. Fiore, North-Holland] 

\noindent Levine, A. M. et al. 1996, ApJ, 469, L33

\noindent Levine, A. M., Morgan, E. \& Muno, M. 1998, IAU Circ. 6966

\noindent Marshall, F. E. 1998, IAU Circ. 6876

\noindent Marshall, F. E., Gotthelf, E. V., Zhang, W., Middleditch, J. \& Wang, Q. D. 
1998, ApJ, 499, L179

\noindent Mendez, M., van der Klis, M., Ford, E. C., Wijnands, R. \& van Paradijs, J. 1999 ApJ 
Letters (in press)

\noindent Mendez, M., van der Klis, M. \& van Paradijs 1998, ApJ, 506,
L117

\noindent Mendez, M., van der Klis, M., Wijnands, R., Ford, E., van Paradijs, J. 
\& Vaughan, B. 1998, ApJ.  505 , L23.

\noindent \noindent Miller, M. C., Lamb, F. K. \& Psaltis, D. 1998, ApJ, 508, 791

\noindent Mirabel, I. F. \& Rodriguez, L.F. 1994, Nature, 371, 46

\noindent Mirabel, I. F.,  Dhawan, V., Chaty, S., Rodriguez, L. F., Marti, J., Robinson, 
C. R., Swank, J. H. \& Geballe, T. 1998, A\&A, 330, L9

\noindent Morgan, E. H., Remillard, R. A. \& Greiner, J. 1997, ApJ, 482, 993

\noindent Nandra, K. et al. 1998, ApJ, 505, 594

\noindent Nowak, M. A., Wagoner, R. V., Begelman, M. C. \& Lehr, D. E. 1997, ApJ, 
477, L91

\noindent Orosz, J. A. \& Bailyn, C. D. 1997, ApJ, 477, 876

\noindent Orosz, J. A., Remillard, R. A., Bailyn, C. D., McClintock, J. E. 1997, ApJ, 
478, L83

\noindent Pooley, G. G. \& Fender,  R. P. 1997, MNRAS, 292, 925

\noindent Psaltis, D. et al. 1998, ApJ, 501, L95

\noindent Radhakrishnan, V. \& Srinivasan, G. 1982, Curr. Sci. 51, 1096

\noindent Remillard, R. A., Wood, A., Smith, D. \& Levine, A. 1997, IAU Circ. 6726; 
see also IAU Circ. 6728

\noindent Remillard, R. A., Morgan, E. M., McClintock, J. E., Bailyn, C. D. \& Orosz, J. 
A. 1999, ApJ, in press (astro-ph 9806049)

\noindent Rothschild, R. E. et al. 1998, ApJ, 496, 538

\noindent Rephaeli, Y. \& Gruber, D. 1999, in preparation

\noindent Rupen, M. P. \& Hjellming, R. M. 1998, IAU Circ. 6938

\noindent Shirey, R. E., Bradt, H. V. \& Levine, A. M. 1999, ApJ (in press) 

\noindent Smith, D. A. et al. 1999, ApJ, in preparation

\noindent Stella, L. \& Vietri, M. 1998, ApJ, 492, L59

\noindent Stella, L. \& Vietri, M. 1999, PRL, 82, 17

\noindent Strohmayer, T. E., Zhang, W., Swank, J. H., Smale, A., Titarchuk, L., Day, 
C. \& Lee, U. 1996, ApJ, 469, L9

\noindent Strohmayer, T., Zhang, W., Swank, J. \& Lapidus, I. 1998, ApJ, 503, L147

\noindent Swank, J. H., Chen, X., Markwardt, C. \& Taam, R. 1997, in Proceedings of 
ÔÔAccretion Processes in Astrophysics: Some Like it HotÕÕ, U. of Md. Oct. 1997, eds. S. 
Holt \& T. Kallman;  astro-ph 9801220

\noindent Taam, R. E., Chen, X. \& Swank, J. H. 1997, ApJ, 485, L83

\noindent Titarchuk, L., Lapidus, I., Muslimov, A. 1998, ApJ, submitted, astro-ph 
9712348

\noindent Tennant, A. F., Fabian, A. C. \& Shafer, R. A. 1986, MNRAS, 221, 27p

\noindent Tingay, S. J. et al. 1995, Nature, 374, 141

\noindent Valinia, A. \& Marshall, F. E. 1998, ApJ, 505, 134

\noindent van der Klis, M. et al. 1996, ApJ, 469, L1

\noindent van der Klis, M. et al. 1997, ApJ, 481, L97

\noindent van der Klis, M. 1998, Nucl. Phys. B (Proc. Suppl.) 69/1--3(1998)103. [Proc. of ÔÔThe Active X-ray SkyÕÕ, eds. L. Scarsi,
H. Bradt, P. Giommi \& F. Fiore, North-Holland]. 

\noindent van der Klis, M. 1999, Proceedings of the Third William
Fairbank, Rome, June 1998; astro-ph 9812395. 

\noindent van Paradijs, J. 1995, in X-ray Binaries, eds. W. H. G. Lewin, J. van Paradijs 
\& E. P. J. van den Heuvel (Cambridge: Cambridge Univ. Press), 536

\noindent Weaver, K. A., Krolik, J. H. \& Pier, E. A. 1998, ApJ, 498, 213

\noindent Wijnands, R. \& van der Klis, M. 1998, Nature, 394, 346

\noindent Zhang, S. N., Cui, W., Chen, W. 1997, ApJ, 482, L155

\noindent Zhang, W. et al. 1998a, ApJ, 495, L9

\noindent Zhang, W., Smale, A., Strohmayer, T. \& Swank, J. 1998b, ApJ. 500, L171

\section*{Figure Captions}

\noindent Fig. 1. Sample of All-Sky Monitor light curves from Mar. 1996 to June 1998 showing, 
top to bottom, a microquasar, the flare star CI Cam, two black-hole binaries, probable disk 
precession in a neutron-star binary, and two faint BL Lac objects. The ordinate is count rate 
adjusted to the center of the field of view of a single ASM camera; the Crab nebula 
would yield $\sim$75 c/s. (A. Levine, pvt. comm.)

\vspace{3 mm}

\noindent Fig. 2. Power density spectra of 4U 1728--34 in three intensity states. Low frequency 
QPO are evident at 20--40 Hz as are two peaks at $\sim$1 kHz which move to higher 
frequencies as the intensity increases. (From Strohmayer et al. 1996)

\vspace{3 mm}

\noindent Fig. 3. Frequency of the two QPO's at kHz frequencies in 4U 1820--30 as a function of 
intensity. The saturation suggests that the innermost stable orbit has been reached, but this conclusion has been questioned --- see text. (Zhang et al. 1998)

\vspace{3 mm}

\noindent Fig. 4. Constraints on mass and radius of neutron star in the non-rotating approximation. 
The highest frequencies detected limit the neutron star mass to $\sim$1.8 $M_{\odot}$. If 
rotation is taken into account, the limits will increase up to at most 2.2 $M_{\odot}$. 
(Miller, Lamb \& Psaltis 1998)

\vspace{3 mm}

\noindent Fig. 5. Dynamic power spectra of two bursts from 4U 1728--34 separated in time by 1.6 
y. In each case, the frequency settles to a stable period at 364.0 Hz with frequencies that 
agree within 0.03 Hz. (From Strohmayer et al. 1998)
\vspace{3 mm}

\noindent Fig. 6. Doppler curve for the 401--Hz pulsar discovered with RXTE. The 
maximum delay is 63 ms, and the binary orbital period is 2.01 h. (From Chakrabarty \& Morgan 1998)
\vspace{3 mm}

\noindent Fig. 7. RXTE/ASM light curve of GRS 1915+105 with hardness ratio (5--12)/(3--5) from Mar. 1996 through Sept. 1998. The marks at the top indicate the times of PCA pointings. (R. Remillard, pvt. comm.)
\vspace{3 mm}

\noindent Fig. 8. Three types of variability of GRS 1915+105 in RXTE/PCA data (E. Morgan, pvt. comm.)
\vspace{3 mm}

\noindent Fig. 9. Precursor outburst activity in GRO 1655--40. The source intensity is shown in the 
optical (BVRI bands) and in the delayed x-ray flux. The onset times are progressively later and 
later as the radiation band hardens. (Orosz et al. 1997)
\vspace{3 mm}

\noindent Fig. 10. Large x-ray dip with spike with simultaneous radio and infrared flares in GRS 
1915+105. Other x-ray dips of this type are shown in the bottom panel of Fig. 8. (Mirabel et al. 1998)
\vspace{3 mm}

\noindent Fig. 11. X-ray character of GRS 1915+105 during and near a dip+spike event, from  RXTE data. Top to bottom: x-ray light curve, inner disk temperature, inner-disk radius, 
photon index of power-law component, dynamic power spectrum. (Swank et
al. 1998)
\vspace{3 mm}

\noindent Fig. 12. Power density spectra of two microquasars showing the high frequency and 
apparently stable QPOs at 67 Hz and 300 Hz. (Morgan, et al. 1997,
Remillard et al. 1999)

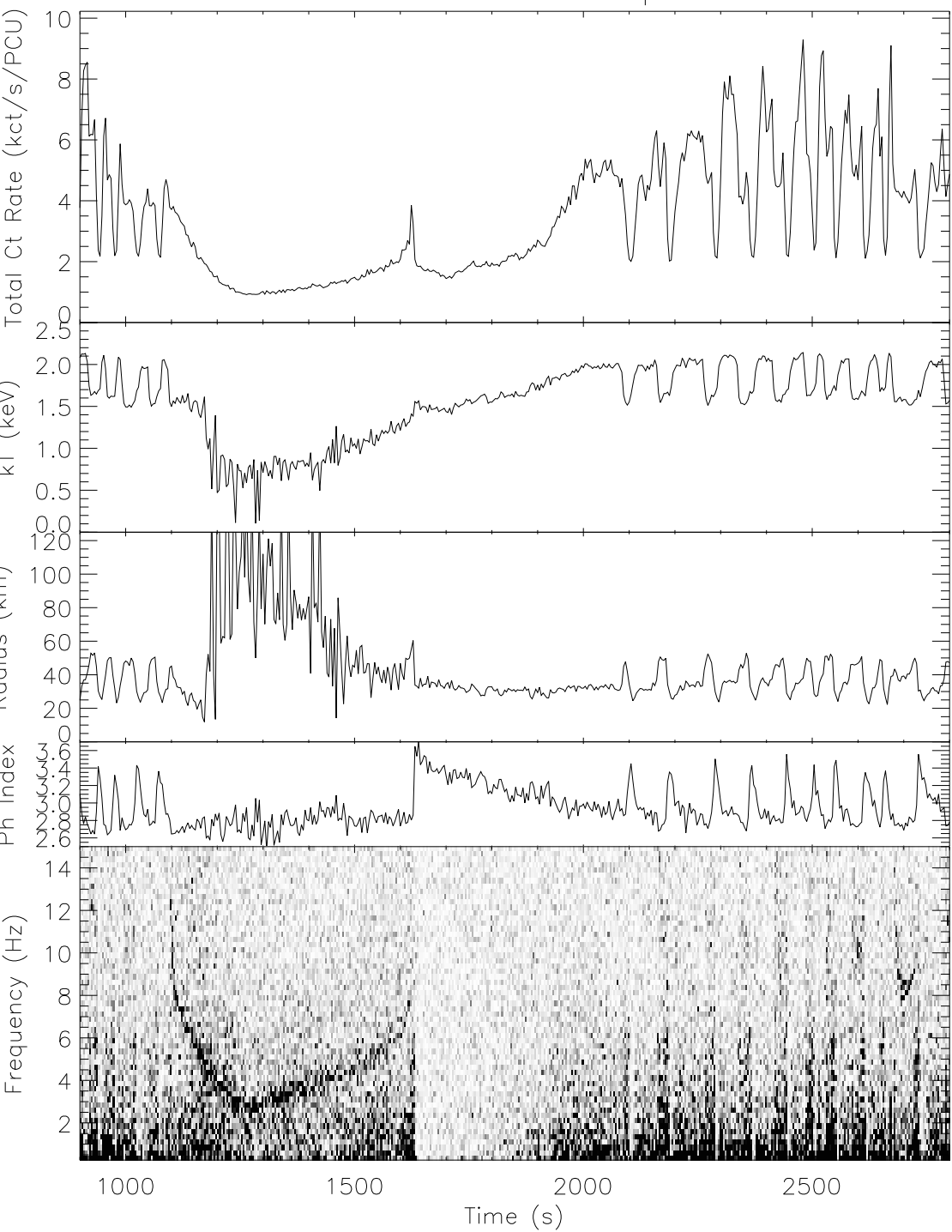
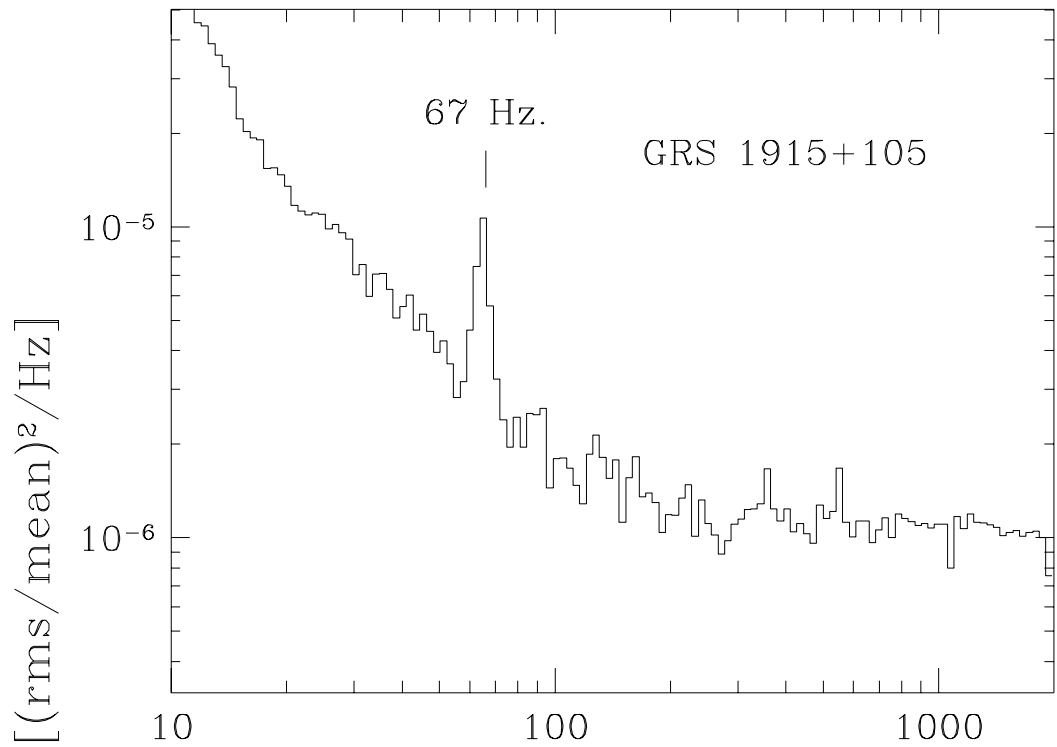
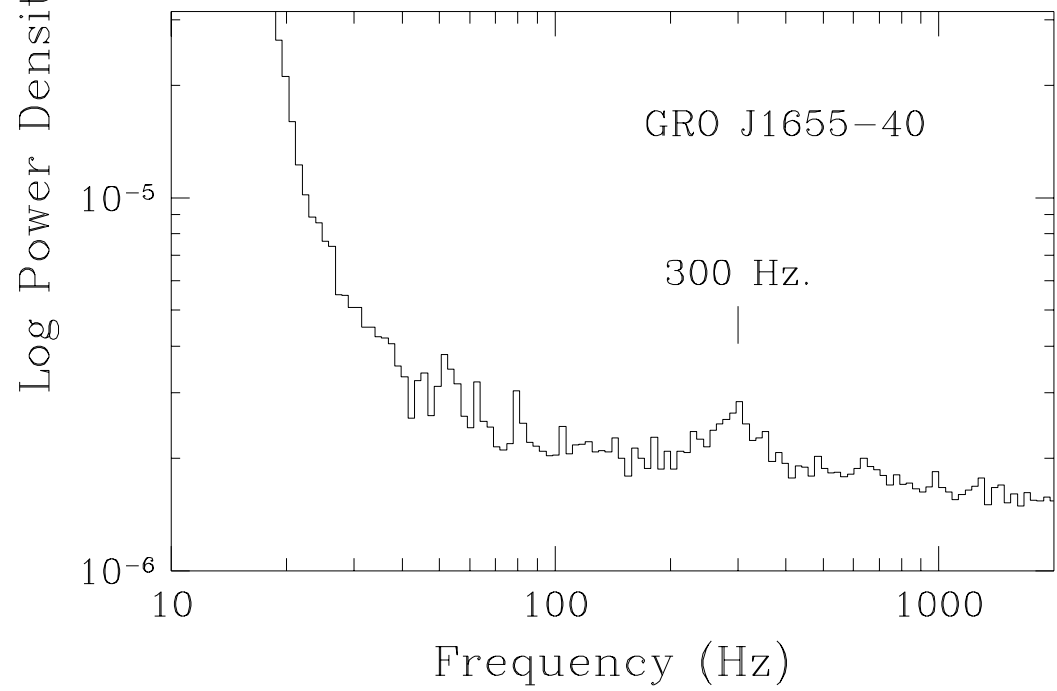
\end{document}